\documentclass{article}

\usepackage{a4,graphicx,amsfonts}

\begin{document}

{\centerline \title {\Large{Study of sediment transport in the saltation regime}}}

{\centerline \large{F. Osanloo$^1$, M. R. Kolahchi$^1$, S. McNamara $^2$, H. J. Herrmann$^3$}}

{\centerline { 1. Institute for Advanced Studies in Basic
Sciences, P. O. Box 45195-1159, Zanjan, Iran}} {\centerline {2.
Institute for Computer Physics, Stuttgart University, 70569
Stuttgart, Germany}}
 {\centerline {3. Computational Physics, IfB,
ETH  Z\"urich, H\"onggerberg, 8093 Z\"urich, Switzerland}}

\begin{abstract}
\end{abstract}

We present a study of sediment transport in the creeping and saltation regime. In our model, a bed
of particles is simulated with
the conventional event-driven method. The particles are considered as
 hard disks in a 2d domain,
with periodic boundary conditions in horizontal direction. The flow of the fluid over this bed of particles is
modeled
by imposing a force on each particle that depends on the velocity of the fluid and its height above the bed.
 We considered two velocity profiles for the fluid, parabolic and logarithmic. The
first one models
 laminar flow and the second corresponds to turbulent flow. For each case we investigated the behavior of
the saturated flux. We found that for
 the logarithmic profile, the saturated flux shows a quadratic increase with the strength of the flow, and for
  parabolic profile, a  cubic increase.
The velocity distribution functions are used to interpret the results.

\clearpage

\section{Introduction}
The study of transport of granular material by a fluid
is important for industrial processes as well as
understanding of natural phenomena. Modeling of the sediment transport in rivers,
as well as modeling of sand drift
in the formation of dunes,
 benefits from this study.

Saltation, surface creep, and suspension are three modes which
occur during transportation of granular material by a
fluid\cite{Bagnold}. When the shear velocity of the fluid flowing
over a bed of grains exceeds the friction threshold velocity for
sand transport, the grains are driven by the fluid. At first they
 begin to move while still in continual
contact with each other,  yet it could happen that every now and
then due to collisions, some particles jump by a distance of order
of their diameter. This regime is called surface creep or
reptation. As the fluid shear velocity increases, the particles
can follow paths that take them to a height much larger than the
grain diameter;
 this regime  is called
saltation. The grains in saltation have been named saltons, and
the grains in creeping motion have been named
reptons \cite{Andreotti1}.
Suspension occurs at very high shear velocities, when a
considerable fraction of the particles are transported upwards by
turbulent eddies. In this regime, the grains move in the fluid
 for long periods of time, hardly colliding with the bed or each other.
 Except for dust storms in which
  suspension is dominant,  creeping  and saltation
  of sands usually play
 the key role in dune formation \cite{You-He Zhou}.
 In many of the experimental studies of sediment transport, grains
are transported by air\cite{You-He Zhou},\cite{Nishimura},  but few
experiments in water also exist\cite{Ancey1},\cite{Ancey2}.

To gain insight into the problem of sediment transport, it is
important to understand the relation between the flux of the
grains transported by the fluid and the velocity profile of the
fluid that moves over the grains. In most instances of sediment
transport, the flux eventually saturates at a certain
strength or amplitude, $u_*$, of the velocity profile, $v(y)=
{u_*} f(y)$. Here, $f$ is a function of height, $y$.
However, there are situations, as at the foot of a sand dune,
where the sand flux may never reach saturation. In such cases, where there is no saturation the
variation of flux can be studied
\cite{Sauermann}.

Bagnold \cite{Bagnold} was first to introduce a simple flux law, a
cubic relation, expressing the behavior of sand flux with the
shear velocity. Other forms for the flux law have also been
proposed by different authors for different physical
conditions\cite{Lettau},\cite{Sorensen}.

In this work, we study two profiles for the fluid flow, and concentrate on the saltation regime
of the grain motion. We emphasize that for the purposes
of this study, saltation has a different meaning than its standard usage.
For instance, in a wind tunnel, a grain is in saltation if its trajectory is
at least about 300 grain diameters high and at least 1000 grain diameters
across. These are much larger than the size of the system
considered here. Yet, we use this term to distinguish the motion from the
situation where the grains constantly touch each other as they
slowly move. Saltation is then used to mean a motion where the grains
jump and follow a trajectory, albeit smaller than  mentioned above.

The structure of the paper is as follows. In Section II we introduce the model.
Then in Section III, we study the behavior of the flux as a function of the velocity profile, as well
as the velocity distribution functions for the grains.
Section IV is devoted to our discussions which mainly rest upon the comparison of velocity
distributions. Finally, we present our conclusions as Section V.

\section{Simulation Model}

We use the inelastic hard sphere model\cite{sean}. Grains are contained in a
two dimensional rectangular domain with
periodic boundary conditions in the horizontal directions.
The fluid flows over the bed of grains, so grains can be entrained
by the fluid.
The fluid is modeled by its velocity
profile.

The grains, modeled as disks, move  under the influence of both
gravity and a drag force that is exerted on them by the fluid. To
avoid crystallizations, $20\%$ of the particles have a diameter
equal to $0.6l_0$ and the rest have a diameter equal to $0.5l_0$
where $l_0$ is the unit of length used throughout  this paper. A
gravitational acceleration of $12l_0/{t_0^2}$ is
applied to all the particles, where $t_0$ is the unit of time  \cite{l0}.

All the particles have the same mass, and the effect of particles on the fluid
is neglected.

\begin{figure}
\begin{center}
\includegraphics[width=2.5in,angle=0]{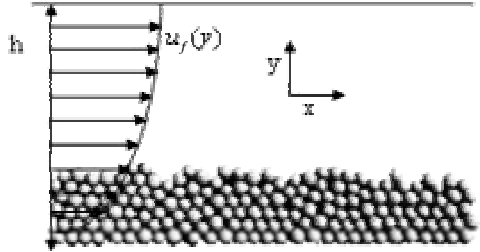}
\end{center}
\caption{The schematic representation of the system showing the
particles in their initial state, and the velocity field at
different positions in the y-direction.} \label{initial}
\end{figure}

In all cases studied, the system starts out having six layers of particles resting fairly compactly
on each other. The fluid stands at a height that is equivalent to thirty two layers, 16$l_0$.
This height is the maximum attainable by the grains; because in our modellization we implement a reflecting
boundary at the top so that the particles that touch it just
reverse their vertical velocity component Fig.(\ref{initial}).

\subsection{ Particle Motion}

We use event-driven MD \cite{event} to calculate the motion of
the particles.
As the particles are hard spheres, collisions take infinitesimal time
and involve only two particles. Conservation of momentum leads to
\begin{eqnarray}
\vec u_{1,2}&=&\vec v_{1,2} \mp \frac{1+r}{2}\left( \hat K\cdot(\vec v_1
-\vec v_2)\right) \hat K \nonumber\\
\label{eq.2}
&\mp& \frac{1+\beta}{2}\left( \hat t\cdot(\vec v_1-\vec v_2)\right) \hat t,
\end{eqnarray}
where $\vec u$ indicates the velocities after the collision, and
$\vec v$ denotes the velocities before the collision. The geometry
of the collision is described by $\hat K$, a unit vector pointing
along the line connecting the centers of particle $1$ to
particle $2$, and $\hat t$ is the unit vector in tangential
 direction. The energy dissipation is
 measured by  $r$, the normal restitution coefficient, and $\beta$, the tangential restitution
 coefficient. If  $r=1$ and $\beta=\pm 1$,
 collisions conserve energy and are said to be elastic.
  For $0<r<1$  or $-1<\beta<1$
 energy is dissipated and the collisions are inelastic.
 We assume that the particles neither rotate nor roll. This is somewhat
 in accord with the fact that the sand grains are not round
 which makes the rolling difficult. In our simulations, $r=0.4$ and $\beta=-1$.

In using the event-driven method there are two problems in setting
up the bed of grains: inelastic collapse, and creating a rough
surface on the bottom. All particles after some collisions lose
their energy and accumulate on the bottom and make a dense network
of grains. So the number of collisions per unit time will diverge
at finite time; that is, inelastic collapse \cite{inelastic}
occurs. Because of
 the finite precision of the computer,
multiparticle collisions can occur. For handling the inelastic
collapse we use the Tc model with $t_c=10^{-6} t_0$
\cite{inelastic}.

 In order to create a rough surface, $r$ and $\beta$ are adjusted for collisions
  between the grains and the surface.   We suppose that when a particle of the first layer
  bounces against the
  bottom plate both the tangential and normal components of its
 velocity are reversed, i.e, $r=1$ and $\beta=1$ in
  Eq.(\ref{eq.2}).  In this way,  the first layer is nearly fixed and acts as a  rough
  surface  over which other particles can move. The roughness is
 of the order of the particle diameter.

\subsection{ The Effect of Fluid on the Grains}

The drag force is proportional to the difference between the
particle velocity $\vec u_p$ and the fluid velocity $\vec u_f$.
\begin{eqnarray}
&& \vec F= \gamma (\vec u_{f}-\vec u_{p}),
\end{eqnarray}
 where $\gamma $ is a parameter that depends on the characteristics
 of both the fluid and the grains. In laminar
 flow with small Reynolds
 number, $\gamma = 3 \pi \eta d_p$, in which $\eta$ is the
 viscosity of the fluid and $d_p$ is the diameter of
  the particle. We suppose that $\gamma=1$ for laminar flow. However in turbulent flow,  the fluid drag varies as
  the square of the grain speed and
  $\gamma$ can be written as:

\begin{equation}
\gamma = \frac{3C_D \rho_{f}}{4\rho_{p}d_p}\mid{\vec u_{f}-\vec
u_{p}}\mid, \label{drag}
\end{equation}
which corresponds to the Newtonian drag force where, $C_D$ is taken from empirical relations,  $\rho_{f}$  and
$\rho_{p}$ are the density of the fluid and particle respectively.

In general, the drag force acts on upper layers of the bed of
particles and drops to zero for lower layers. The details depend
on the velocity profile considered. Here, we study the dynamics of
the grains for two velocity profiles: logarithmic and parabolic.
For the parabolic  profile, which models laminar flow in open
channels as function of height,
 we have:
\begin{equation}
u_f = {u_*}(y_0(y_0/2-h)-y(y/2-h)),
\label{e4}
\end{equation}
where  $h$ is the height of fluid in channel. This equation is written so that
it satisfies the two boundary conditions, $\partial u_f / \partial y =0$ at $y=h$, and $u_f=0$ at
$y=y_0$. Here, $y$ is the height, and $y_0$ is the height below which
the effect of the fluid on the grains is negligible.

In turbulent flow, the  velocity profile of the fluid near the
boundary is observed to be logarithmic\cite{alan}, and described
by
\begin{equation}
u_f = {\frac{u_*}{k}}log(y/y_0 ),
\label{e5}
\end{equation}
where $k$ is the von Karman constant. Although this relationship
has been derived only for the region in which the shear stress is
approximately constant, experiments show that the agreement
persists through almost all of the boundary layer. In our
simulations, $y_0$ is about the diameter of grains, $y_0=0.5 l_0$.
We consider no vertical component to the fluid velocity.

\section{Granular Flux and Velocity Probability Distribution Function}
First we investigate the behavior of the granular flux with
respect to time. Granular flux is the number of grains that cross
the unit surface (unit line in 2d ) per time unit. So the flux
integrated over a cross section of the periodic domain (over
height in 2d) is:
\begin{equation}
q=\frac{1}{L}\sum_{i=1}^{N}u_{i},
\label{e6}
\end{equation}
where  $L$ is the length of the box and $N$ is the total number of
particles, and $u_i$ denotes the horizontal component of the
velocity of the $i$th   particle.

After some time the flux fluctuates around a constant, steady value, and sediment transport reaches
steady state. This steady state is the saturated flux, and it depends
on the shear velocity, denoted by $u_*$.

For calculating the saturated flux value, we average over the flux only
after the steady state has been reached.
We expect to have a threshold velocity $u_t$, below which sediment transport cannot happen.

\subsection{Logarithmic Profile }

\begin{figure}
\begin{center}
\includegraphics[width=3.5in,angle=0]{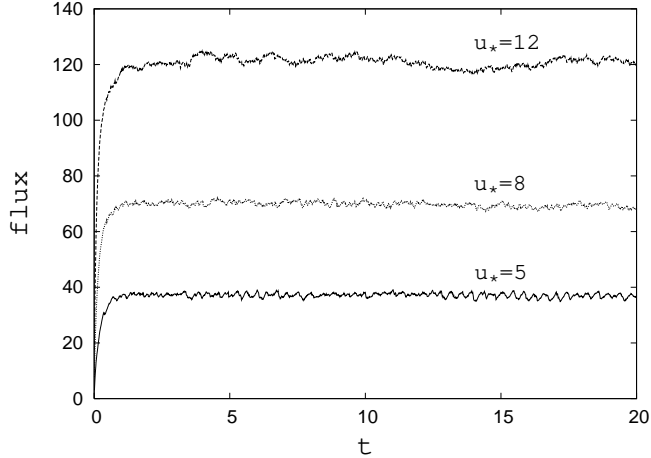}
\end{center}
\caption{Flux as a function of time for different values of $u_*$ for a logarithmic profile.}
\label{f-t}
\end{figure}

In the aeolian case the logarithmic velocity profile is more
realistic than the parabolic profile. When the wind  blows over a
rough surface, its velocity within the boundary layer increases
logarithmically with height, Eq.(\ref{e5}). The logarithmic
profile has also
 been observed experimentally,
for water moving over a rough surface in a channel \cite{Ancey1, Ancey2}.

 Fig.(\ref{f-t}) shows the variation of  flux with time for different values of shear velocity, $u_*$.
 This figure shows that transport process reaches steady state and the particle
 flux saturates after some transient time.

 The simulations could be made into movies of the grain motion. This was a
 particularly useful way of interpreting the results.
 In this way we estimate the grain motion to start at a threshold velocity of about $u_*=1$.
 With increasing $u_*$ from $u_t=1$, the top layer particles begin
 to roll in their own layer or jump to a height about their diameter. This situation continues
 until $u_* \simeq 15$. This means that for $u_*<15$ most of the particles
 except those in the bottom layer are in the creeping regime.
 After $u_*>15$ the shear stress is enough to make some of the particles
 in the upper layers enter the saltation regime. With increasing
 $u_*$, the number of
 saltating particles increases, resulting in an increase in the number of collisions between saltons.
 We estimate $u_* \simeq 15$ as marking the onset of
  saltation in this system. When $u_*>30$ enough particles have so high an energy that
 they move with the fluid stream above the other particles and have few collisions with each
 other or the rest of the grains.
 In this case the length of their trajectory
 becomes comparable to the system size, hence we only
 considered $u_*<30$.  We wish to emphasize again that we are using
 the term saltation in a restricted
  sense in this study.

\begin{figure}
\begin{center}
\includegraphics[width=3.5in,angle=0]{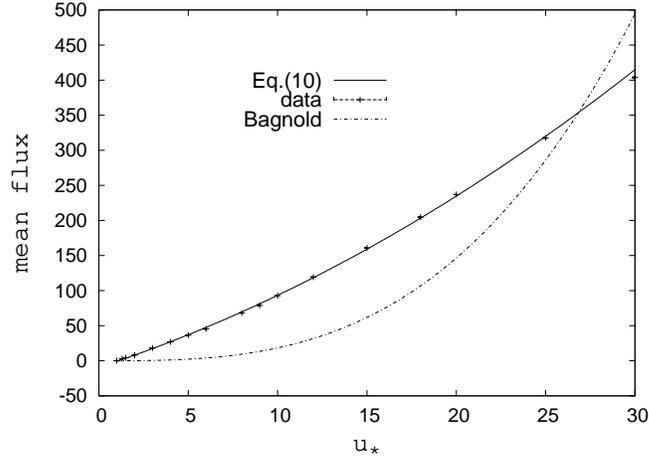}
\end{center}
\caption{Mean flux vs.  $u_*$ for a logarithmic profile, and the
fit to it, Eq. (\ref{e10}) and Bagnold's expression.
Fit the Eq. (\ref{e10}) gives $a=0.2$, $b=43.9$, and $u_t=1.09$.
According to the simulations we estimate $u_t=1$. }
\label{mean-f}
\end{figure}

The main objective of this study is to relate the saturated flux and
the shear velocity in the regime of saltation. Our results for
the behavior of the mean flux of particles with $u_*$ is shown in Fig.(\ref{mean-f}).

Theoretically, perhaps the most important description for saturated flux is due to Bagnold \cite{Bagnold}.
He found
that the saturated flux at large shear velocities is given by:
\begin{equation}
q = \frac{\rho_{air}}{ g} u_*^3.\label{e7}
\end{equation}

Bagnold considered a mean trajectory for each grain, and supposed that the ejection velocity of
 grains from the bed scales with the shear velocity $u_*$. This hypothesis is valid if the
shear velocity is large enough. For small shear velocities, Ungar \& Haff \cite{Bruno} supposed that
height and length of
 the trajectory of grains is of the same order as the grain size, and predicted that:
 \begin{equation}
q \propto \rho_{air}(u_*^2-u_t^2) \sqrt{\frac{d}{g}},\label{e8}
\end{equation}
where $d$ is the grain diameter. Almeida et {\it al.}\cite{Almeida} also found numerically a
quadratic description for
 the flux near the threshold shear velocity,
 \begin{equation}
q \propto (u_*-u_t)^2\label{e9}
\end{equation}

They simulated the saltation inside a two-dimensional channel with
a mobile top wall. Their model solves the turbulent wind field
including  feedback of the dragged particles.

Our results show a slightly different quadratic dependence:
\begin{equation}
q=a(u_*-u_t)(u_*+b)\label{e10}
\end{equation}
where $a=0.2$, $b=43.9$, and $u_t=1.09$ are obtained from fitting the Eq.(\ref{e10}) to our
data. This value for the threshold velocity is in good agreement
with what was estimated from the simulations; that is $u_t=1$.
Fig.(\ref{mean-f}) shows the fit, Eq. (\ref{e10}), to the data.
 This indicates that both in the creeping regime and the
saltating regime, the same quadratic function describes the
behavior of flux reasonably well. Eq. (\ref{e10})  predicts a
stronger dependence on shear velocity, compared to Eq.
(\ref{e9}). One reason may  be our neglect of the back action of
the grains on the fluid. This effect is more prominent at small
heights above the bed where
the particle velocity differs much from the fluid velocity
\cite{Almeida}.

\begin{figure}
\begin{center}
\includegraphics[width=3.5in,angle=0]{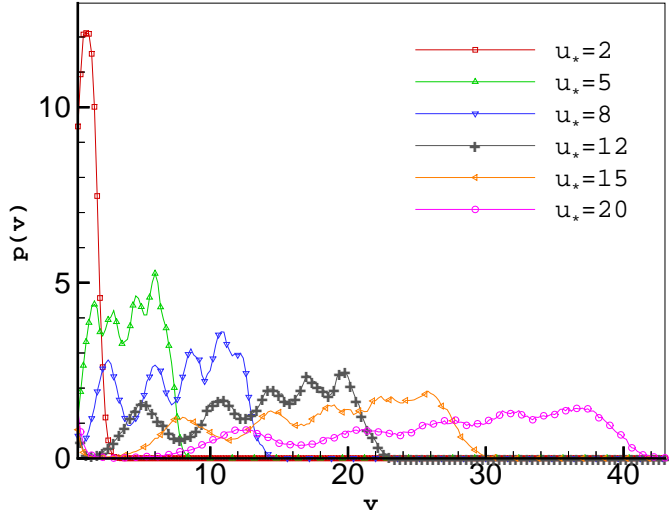}
\end{center}
\caption{The behavior of grain velocity probability function for
different values of $u_*$ for the logarithmic profile.}
\label{pdf1}
\end{figure}

Another way to estimate the onset of saltation is to
investigate the behavior of the grain velocity probability
distribution function. The distribution function allows
decomposition of the flux into a part due to saltation and another
part due to creep. This interpretation is based on the fact that
the reptons are slower than the saltons. Fig.(\ref{pdf1}) shows
the behavior of the grain velocity distribution for different
values of $u_*$.

For small values of $u_*$ there is a large peak at small
velocities that shows that all of the particles are in creeping
motion. With increasing $u_*$, the  velocity of the moving
particles increases and some of the particles enter
the saltation regime. The uniform distribution function  at
higher velocities  represents the saltating particles.  At  $u_*=15$ the particles
can go to the saltation regime; the velocity distribution develops
a minimum at $v \sim 1$.
 At  $u_* \sim 20$ all of particles have saltating motion.

\subsection{Parabolic Profile }

At low Reynolds numbers when a fluid flows in an open channel ,
its velocity varies with height parabolically. Now, we consider a
velocity profile as in Eq.(\ref{e4}), that is zero
at the bottom and increases parabolically
 with height.

\begin{figure}
\begin{center}
\includegraphics[width=3.5in,angle=0]{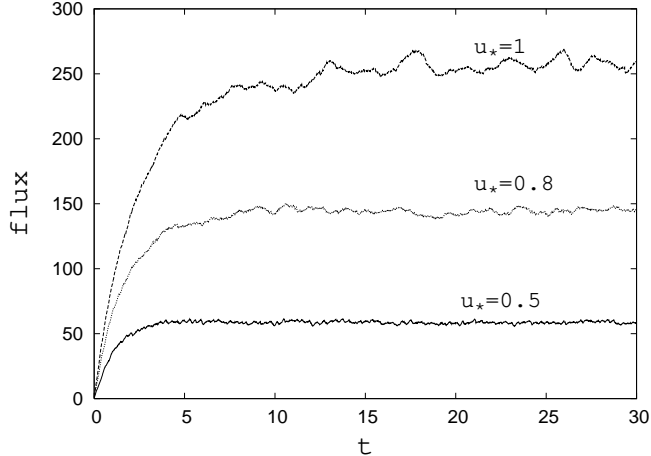}
\end{center}
\caption{Flux as a function of time a for parabolic profile.}
\label{f-t3}
\end{figure}

Fig.(\ref{f-t3}) shows the behavior of flux as a function of time. Similar to
 the logarithmic profile, the flux
reaches steady state, only now  the transient time is longer.
In the parabolic profile, the shear stress above the bed is greater than the logarithmic profile and increases
rapidly with $u_*$,
so compared to the logarithmic case,
 the range of
$u_*$ that defines the creep motion is much narrower.

The motion starts out at about $u_t=0.08$ \cite{ut}; this is an estimate obtained
from our simulations, and in Eq.(\ref{e12}) below is used as an input data. When $u_*=1$ the system is out of the creeping regime,
and starts saltating.
The surface $y=h$ is as before a reflecting surface, so any particle that reaches it
 bounces back into the fluid. This starts at about
$u_*=1.2$.

\begin{figure}
\begin{center}
\includegraphics[width=3.5in,angle=0]{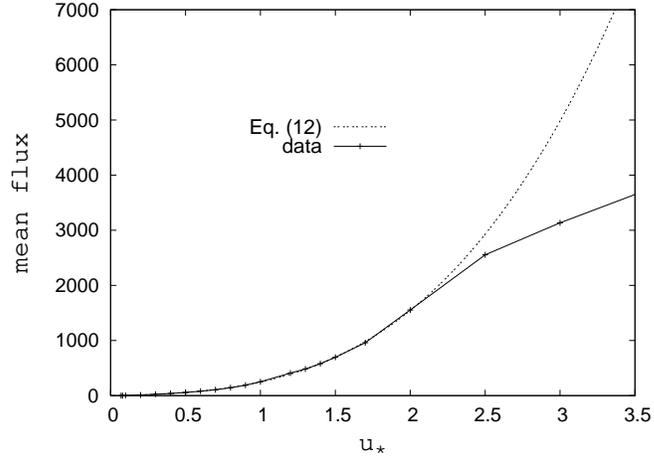}
\end{center}
\caption{Mean flux as a function of $u_*$ for a parabolic
profile. The solid line is our data, and the dashed line is fitting a cubic function to data. Beyond $u_*
\sim 2.5$ the effect of the `top boundary` is revealed. $u_t=0.08$ is estimated from the simulations.}
\label{f6}
\end{figure}

Beyond $u_*=2.5$ this boundary produces a kind of population
inversion  and the collision of particles with each other effectively reduces the rate
of increase of flux Fig. (\ref{f6}).

\begin{figure}
\begin{center}
\includegraphics[width=3.5in,angle=0]{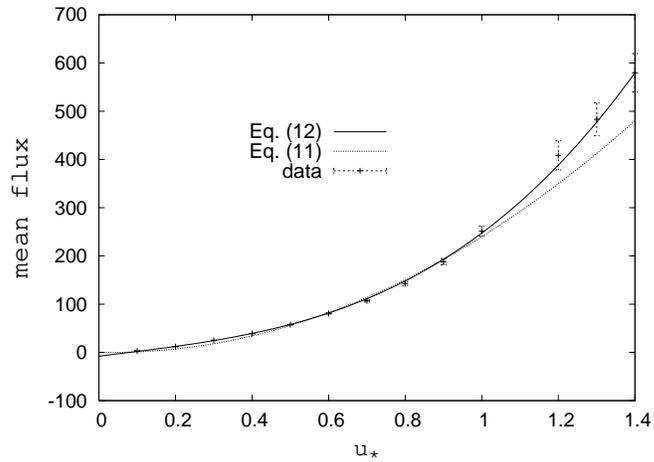}
\end{center}
\caption{Mean flux as a function of $u_*$ near the threshold value $u_t=0.08$ for a parabolic profile. For  $u_*<1$ we fitted
the data with two
 Eqs.(\ref{e11}) , (\ref{e12}). The solid line is the fit using Eq. (\ref{e12}) yielding $a=182$,
 $b=-0.07$,
and $c=0.5$. The dashed line is the fit using Eq. (\ref{e11})
that gives $a'=258$, and $u_t=0.036$.} \label{linear}
\end{figure}

In fig.(\ref{linear}) we show the behavior of the mean flux with
$u_*$ (near $u_t$). For  $u_*<1$ we fitted the data with two
equations below:
\begin{eqnarray}
q&=&a'(u_*-u_t)^2\label{e11}\\
q&=&a(u_*-0.08)(u_*^2+bu_*+c)\label{e12}
\end{eqnarray}
The values of  $a'=258$, $a=182$, $b=-0.07$, $c=0.5$, and $u_t= 0.036$ are obtained from fitting.
The cubic equation is clearly a better fit to the data than the
quadratic one.

\begin{figure}
\begin{center}
\includegraphics[width=3.5in,angle=0]{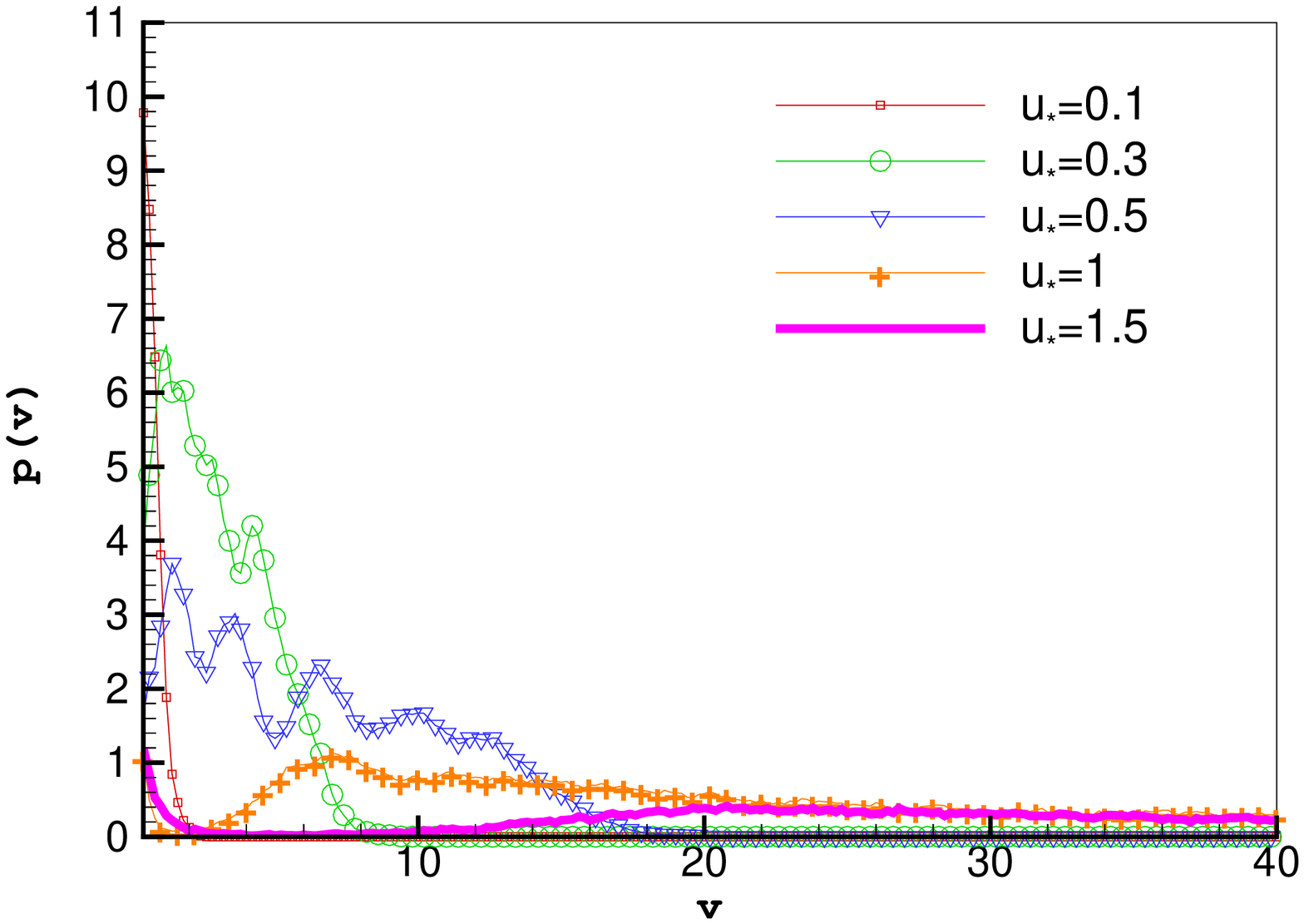}
\end{center}
\caption{The behavior of the grain velocity probability function
for different values of $u_*$ for a parabolic profile.}
\label{pdfp1}
\end{figure}

Fig.(\ref{pdfp1}) shows the behavior of the grain velocity
distribution for the parabolic profile. In this case, the
saltation regime starts at $u_*=1$, below which
 all particles are
in creeping motion. Beyond $u_*=1.5$, all particles are in the
saltation regime.

\section {Discussion}

 The dynamics of the granular systems under shear is usually
 studied for slow or quasi static flow \cite{gollub}.
 Slow shearing can lead to the emergence of internal ordering
 of the granular systems. In terms of our study, this means
 the creeping regime. In fact we have also observed size segregation
 for the case of the logarithmic profile when the shear is small enough.
 Two granular materials differing by their size,
 exhibit a propensity for segregation.
 Whenever this mixture undergoes a vibration or a shearing action, the components tend to
 separate partially or completely \cite{segregate}.
Here we considered a system with two different sizes, differing in
ratios by a factor 2, initially  distributed randomly.
When this system is affected by the drag force of
the fluid, the two components of the system separate so that  large
particles will come to the top. In the creeping regime
due to the higher collision rate,  this separation occurs
very soon.

 Overall, the present study considers the intermediate
 region, between the quasi static flow and the rapid flow.
  Due to finite size effects, we use the term `saltation',
 to mean a motion that takes a particle on a trajectory, however small,
 raises it from the bed of grains.

 We interpret the results using the grain velocity distribution function, $P(v)$.
 With increasing shear we expect to have more and more
 particles in motion. The distribution function, $P(v)$, develops a peak
 at low velocities, $v \sim 2$, that gradually disappears as $u_*$ increases.
 In case of the logarithmic profile, this peak does not broaden.
 With the diminishing of this low energy peak, a plateau is developed
 in the velocity distribution. This plateau gradually covers
 a larger range of velocities. This plateau corresponds to saltons. The density of particles
 in the saltation regime, for the logarithmic profile,
 is a monotonically decreasing function
 of height. We attribute this to the comparatively smaller energy
 input rate; the latter going as $q_f u_f^2$, where $q_f$ is the fluid flux.

 For the parabolic profile, the situation is markedly different.
 We observe that as the shear velocity increases, the  distribution
 develops a tail, covering the larger velocities.
 The rate of energy input is
 much higher in the case of the parabolic profile. In the saltation regime $(u_*\geq 0.8)$,
 it is possible to
 distinguish a group of particles that moves at a higher velocity and height,
  nearly separating from the rest
 of the particles. In this sense, the density of particles in the saltation regime, is
 qualitatively different
 from that of the logarithmic velocity profile. This explains the sudden velocity
 spread at $u_*\geq 0.8$.

\section{ Conclusion}

In this study, we have presented a simple model for sediment transport in creeping
and saltating motion that
produces steady sediment transport. We investigated the steady state
 fluxes for the parabolic and logarithmic
fluid velocity profiles.

For the logarithmic profile we compared our results with some
previous studies. Increasing the shear velocity $u_*$ from
threshold value $u_t$, sediment transport sets in,
 first  creeping  and
for larger $u_*$,  saltating. In our system, the saltating
particles
 can rise up
to several times their diameter, and similar to the Ungar and Haff
we find that the steady state flux increases quadratically with
shear velocity. For the parabolic distribution the flux rises
cubically with increasing the shear velocity.

The velocity probability distribution function is a good measure
to estimate the onset of the saltation. It would be
interesting to study a system large enough so that the grain
velocity distribution function  can be
measured as a function of height.

\vspace{0.5cm}
{\Large{Acknowledgment}}
\vspace{0.5cm}

Part of this work was carried out while F. Osanloo stayed at the University of Stuttgart.
She wishes to thank the University of Stuttgart for their hospitality and the grant that made
her stay possible.

\bibliographystyle{}      

\begin{thebibliography}{}
\bibitem{l0}Throughout this paper the unit of length is $l_0=2*10^{-1}cm$, and the gravitational acceleration
of $12l_0/t_0^2=490 cm/sec^2$ is applied to all the particles. This is the acceleration
experienced by a particle with density of $2g/cm^3$ in water. The corresponding unit of time is
$7*10^{-2} sec$.
\bibitem{ut} Although the threshold velocities for the two profiles are drastically different,
the resulting drag forces (for the top layer), using Eqs. (2) through (5), come out nearly the same for both profiles, as
 expected, at the onset of motion.
\end{thebibliography}

\end{document}